**Article title**

Zipfian universality of interaction laws: A statistical-mechanics framework for inverse power scaling

**Author**

Jérôme Baray[1]

**Affiliation**

[1] Le Mans Université
Laboratoire ARGUMans (EA 4321)
Avenue Olivier Messiaen
72085 Le Mans Cedex 9
France

Email: jerome.baray@univ-lemans.fr

# Zipfian universality of interaction laws: A statistical-mechanics framework for inverse power scaling

## Abstract

Inverse power-law interaction forms, such as the inverse-square law, recur across a wide range of physical, social, and spatial systems. While traditionally derived from specific microscopic mechanisms, the ubiquity of these laws suggests a more general organizing principle. This article proposes a statistical-mechanics framework in which such interaction laws emerge as macroscopic fixed points of aggregation processes involving strongly heterogeneous microscopic contributions.

We consider systems where individual interaction sources exhibit heavy-tailed heterogeneity consistent with Zipf–Pareto statistics and where aggregation proceeds without intrinsic length scales. Under minimal assumptions of heterogeneity, multiplicativity, scale invariance, and stability under coarse-graining, we show that the resulting macroscopic interaction field must adopt a scale-free, power-law form. The associated exponent is not imposed a priori but emerges from effective dimensionality, symmetry, and aggregation structure.

Within this framework, the inverse-square law is interpreted as a stable statistical fixed point corresponding to isotropic aggregation in an effective three-dimensional space, while deviations from this regime naturally arise from anisotropy, constrained geometries, or nontrivial effective dimensions. This perspective provides a unified interpretation of interaction laws observed in physics, spatial economics, and human geography, without invoking domain-specific microscopic mechanisms.

The proposed framework reframes inverse power-law interactions as robust emergent features of Zipfian aggregation rather than as unique consequences of particular physical forces, thereby offering a common statistical explanation for their cross-disciplinary recurrence.

# Zipfian universality of interaction laws: A statistical-mechanics framework for inverse power scaling

**1. Introduction**

Inverse power-law interaction forms are among the most recurrent and robust regularities observed across physical, natl and socio-spatial systems [1]. From Newtonian gravitation and Coulomb electrostatics to radiative intensity, diffusive fluxes, and a wide class of spatial interaction models in geography and economics, effective interactions often decay with distance according to a simple scaling law of the form

$$I(r) \sim r^{\{-\alpha\}}$$

$\alpha$ typically takes low integer values, most prominently $\alpha = 2$. The striking recurrence of such laws across domains governed by radically different microscopic mechanisms has long suggested the presence of a deeper organizing principle, transcending the specific physical, institutional, or behavioral foundations usually invoked in their derivation.

In their standard interpretations, inverse-power interaction laws are derived from domain-specific arguments. In classical physics, inverse-square laws arise naturally from field conservation and geometric spreading in isotropic Euclidean space, as formalized through Gauss's theorem and the structure of Maxwell's and Newton's equations. In diffusion and transport phenomena, power-law decays emerge from continuum limits of local flux balances. In spatial economics and geography, gravitation-like models are motivated by analogies with physical attraction, supplemented by empirical calibration. While these derivations are mathematically sound within their respective frameworks, they remain largely disconnected from one another and offer little explanation for the remarkable universality of the resulting functional forms.

This situation raises a fundamental question that remains insufficiently addressed: *why do inverse power-law interactions emerge so systematically in systems whose microscopic constituents, interaction rules, and governing equations differ so profoundly?* The recurrence of similar scaling exponents across physics, biology, and social systems suggests that inverse-power laws may not be primarily tied to the detailed nature of interactions, but rather to generic statistical mechanisms operating at large scales.

In parallel, a vast body of empirical and theoretical work has documented the ubiquity of Zipf and Pareto laws in complex systems [1–4]. Originally identified in linguistics and later observed in city sizes, firm distributions, income, network connectivity, and many other contexts, Zipfian scaling reflects the emergence of heavy-tailed distributions characterized by scale invariance and the absence of a typical size. Despite their pervasiveness, Zipf and Pareto laws are most often treated as empirical regularities or phenomenological outcomes of specific growth or preferential attachment processes, and a general theoretical explanation of their universality remains an open issue [5,6]. Their potential role as a unifying statistical mechanism underlying macroscopic interaction laws has received comparatively little attention.

This paper proposes a theoretical framework that explicitly connects these two strands of universality. We argue that inverse power-law interaction forms can be understood as emergent statistical fixed points of Zipfian aggregation processes. In this perspective, Zipfian scaling is not merely an empirical descriptor of size or rank distributions, but a generative statistical mechanism arising from the aggregation of heterogeneous microscopic contributions under scale-invariant conditions. When such aggregation processes are spatially mediated, they naturally give rise to effective interaction laws that decay as inverse powers of distance.

The central idea developed here is that many interaction laws commonly regarded as fundamental or domain-specific—such as Newtonian gravitation, Coulomb electrostatics, or spatial gravity models in human systems—can be reinterpreted as particular realizations of a more general Zipfian statistical-mechanics framework. Within this framework, microscopic interaction costs, intensities, or propensities are assumed to be heterogeneous and multiplicative in nature. Their aggregation over space, in the absence of a characteristic scale, leads to heavy-tailed effective fields whose macroscopic form is governed by universality rather than microscopic detail.

From this viewpoint, inverse-power interaction laws emerge as statistical attractors. The exponent $\alpha$ is not postulated *a priori*, nor uniquely dictated by the underlying micro-physics, but results from symmetry, dimensionality, and aggregation constraints. In particular, the inverse-square law appears as a privileged fixed point associated with isotropic aggregation in effectively three-dimensional space, while deviations from this exponent naturally arise when these assumptions are relaxed. This interpretation provides a unified statistical explanation for both the robustness of inverse-square laws and the observed variability of exponents in empirical interaction systems.

The contribution of this paper is threefold. First, we formalize Zipfian aggregation as a general statistical-mechanics process capable of generating long-range interaction laws. Second, we demonstrate that inverse power-law interactions correspond to fixed points of this process, with the inverse-square law emerging as a special symmetric case. Third, we show that well-known interaction models—including Newtonian gravitation, Coulomb electrostatics, and spatial gravity models such as Reilly-type formulations [7]—can be consistently interpreted within this unified framework as particular instantiations rather than isolated principles.

By reframing interaction laws as emergent statistical phenomena rather than fundamental postulates, the proposed framework shifts the focus from domain-specific derivations to universal mechanisms of aggregation and scaling. This approach does not challenge the validity of established physical or spatial laws; instead, it provides a complementary interpretation that clarifies why similar functional forms recur across disciplines. In doing so, it opens the way to systematic generalizations beyond classical inverse-square laws and offers a coherent statistical foundation for the study of interaction phenomena in complex systems.

The remainder of the paper develops this framework in detail. Section 2 introduces the Zipfian aggregation principle and situates it within the broader context of statistical mechanics and universality. Section 3 presents the general formulation of Zipfian interaction laws and derives their macroscopic scaling behavior. Section 4 identifies the inverse-square law as a statistical fixed point and discusses its relation to classical interaction models. Section 5 explores extensions beyond the inverse-square regime, while Section 6 provides a minimal illustrative example. The paper concludes with a discussion of implications and open directions.

## 2. Zipfian aggregation and universality

The purpose of this section is to elevate Zipfian scaling from the status of an empirical regularity to that of a general statistical principle. In doing so, we deliberately focus on statistical mechanisms rather than system-specific models, in order to isolate the universal features underlying Zipfian scaling. Rather than treating Zipf's law as a descriptive outcome tied to specific systems—such as word frequencies, city sizes, or firm distributions—we interpret it as the macroscopic signature of a broad class of aggregation processes characterized by heterogeneity, multiplicativity, and scale invariance. Within this perspective, Zipfian behavior is neither accidental nor domain-specific, but emerges naturally whenever these statistical conditions are met.

### 2.1 Power laws, heavy tails, and scale invariance

Power-law distributions occupy a central place in the study of complex systems [1,5]. A random variable (X) is said to follow a power-law distribution if its tail probability decays as

$$P(X > x) \sim x^{-\beta},$$

for large values of x, with $\beta > 0$.

Such distributions are characterized by heavy tails, implying that extreme events are not exponentially suppressed and that no characteristic scale governs the system. This absence of scale is formally expressed through scale invariance: under a rescaling of the variable, the functional form of the distribution remains unchanged up to a multiplicative constant.

Zipf's law represents a particular manifestation of power-law behavior, typically expressed in rank–frequency form. If elements are ordered by decreasing magnitude, the value associated with rank (k) scales as

$$X(k) \sim k^{-\zeta},$$

with $\zeta \approx 1$ in many empirical systems. Rank-based and size-based formulations are mathematically equivalent under broad conditions, and both reflect the same underlying scale-free structure. Importantly, the relevance of Zipfian scaling lies not in the precise numerical value of the exponent, but in the fact that the distribution lacks a characteristic scale and exhibits strong heterogeneity across orders of magnitude.

Heavy-tailed and scale-invariant distributions contrast sharply with exponential or Gaussian statistics, which are dominated by a typical scale and rapidly suppress large deviations. As a result, systems governed by power laws are structurally dominated by a small number of large contributions, while the majority of elements contribute marginally. This asymmetry plays a crucial role in the emergence of long-range effects and effective interactions at macroscopic scales.

**2.2 Zipfian scaling as an outcome of multiplicative aggregation**

A key insight from statistical physics and applied probability is that power-law distributions often arise from multiplicative, rather than additive, aggregation processes. Consider a quantity constructed as the product of many heterogeneous random factors,

$$X = \prod_{i=1}^{N} Y_i,$$

where the $Y_i$ are independent or weakly dependent positive random variables. Taking logarithms transforms this product into a sum,

$$\log X = \sum_{i=1}^{N} \log Y_i.$$

If the logarithmic increments $\log Y_i$ have sufficiently broad distributions or exhibit nontrivial correlations, the resulting distribution of $X$ can develop heavy tails. In many cases, this mechanism leads to lognormal distributions at intermediate scales and power-law tails in the asymptotic regime, depending on boundary conditions and normalization constraints.

Zipfian scaling can thus be interpreted as the macroscopic outcome of multiplicative growth or aggregation under constraints that prevent indefinite divergence [1,4]. Such constraints may take the form of normalization, competition, conservation laws, or finite resources. Importantly, the microscopic details of the factors $Y_i$ are largely irrelevant: what matters is the multiplicative structure and the absence of a characteristic scale. This explains why Zipf's law appears in systems with widely different microscopic interpretations, ranging from linguistic usage to economic activity and network connectivity.

In this sense, Zipfian scaling is not tied to any specific generative model, such as preferential attachment or proportional growth, but reflects a broader universality class of aggregation processes. Different microscopic mechanisms can belong to the same class and give rise to indistinguishable macroscopic statistics.

**2.3 Comparison with additive aggregation and the central limit theorem**

The contrast between Zipfian aggregation and classical additive aggregation is instructive. In additive processes, a macroscopic quantity is constructed as a sum of many independent contributions,

$$S_N = \sum_{i=1}^{N} X_i.$$

Under very general conditions, the central limit theorem (CLT) ensures that the properly rescaled sum converges to a Gaussian distribution as $N \to \infty$, regardless of the detailed distribution of the individual terms [8]. The Gaussian distribution thus represents a universal fixed point for additive aggregation of finite-variance variables, as commonly discussed in statistical mechanics and critical phenomena.

Zipfian aggregation corresponds to a fundamentally different regime. When aggregation is multiplicative, or when additive contributions themselves exhibit diverging variance or strong correlations, the assumptions of the CLT are violated. In such cases, Gaussian statistics no longer apply, and alternative fixed points emerge. Power-law and Zipfian distributions can be viewed as the analogs of Gaussian distributions for aggregation processes lacking a characteristic scale.

From this perspective, Zipfian scaling plays a role analogous to that of the Gaussian in classical statistics: it represents a universal attractor for a broad class of non-additive, scale-free aggregation processes. This analogy provides a first step toward interpreting Zipf's law as a fundamental statistical principle rather than a system-specific curiosity.

**2.4 Relation to stable laws and Lévy statistics**

The breakdown of the CLT in the presence of heavy-tailed variables leads naturally to the theory of stable distributions. Stable laws generalize the Gaussian distribution to cases where the variance of the underlying variables diverges. Lévy stable distributions are characterized by power-law tails and are themselves invariant under aggregation up to rescaling [9,10].

Zipfian distributions are closely related to this family of stable laws, although they are most naturally expressed in rank space rather than probability density form. Both share the defining property of scale invariance and the dominance of rare but extreme events. However, while stable laws describe the distribution of sums of heavy-tailed variables, Zipfian scaling captures the hierarchical organization of magnitudes resulting from repeated aggregation and normalization.

This distinction is important for interaction systems. In many contexts, the relevant quantity is not the sum of contributions but the effective strength or influence resulting from multiplicative combinations of heterogeneous factors. Zipfian aggregation provides a natural statistical description of such situations, complementing the role played by stable laws in additive settings.

## 2.5 Renormalization and universality

The emergence of Zipfian scaling can also be interpreted through the lens of renormalization. In statistical physics, renormalization-group approaches describe how macroscopic behavior emerges from microscopic interactions through successive coarse-graining transformations. Systems that flow toward the same fixed point under renormalization share identical large-scale properties, regardless of microscopic detail [11–13].

Zipfian distributions can be understood as fixed points of renormalization flows associated with scale-free aggregation. Under coarse-graining, the relative importance of large contributions is preserved, and the absence of a characteristic scale remains invariant. This interpretation reinforces the view that Zipfian scaling defines a universality class, in the same sense as critical phenomena in equilibrium statistical mechanics.

Within this framework, the precise value of scaling exponents depends on symmetry, dimensionality, and constraint structure, but the existence of power-law behavior itself is robust. This observation is central to the argument developed in this paper: inverse power-law interaction forms should be interpreted as manifestations of Zipfian universality at the level of effective interactions. These considerations suggest that Zipfian universality should naturally extend to effective interaction fields emerging from aggregation over space.

## 2.6 Definition of the Zipfian aggregation process

We are now in a position to define the concept of a *Zipfian aggregation process*. We refer to a Zipfian aggregation process as any aggregation mechanism satisfying the following conditions:

1. Heterogeneity: microscopic contributions are broadly distributed and not characterized by a single typical scale.

2. Multiplicativity: macroscopic quantities result from multiplicative or compounding combinations of microscopic factors.

3. Scale invariance: the aggregation process does not introduce a characteristic scale, either through normalization or constraint.

4. Stability under coarse-graining: the macroscopic distribution is preserved under aggregation of aggregated units.

Under these conditions, the resulting macroscopic statistics belong to the Zipf–Pareto family. Importantly, this definition is independent of the specific nature of the microscopic variables and applies equally to physical, biological, economic, and spatial systems.

In the remainder of this paper, we show that spatially mediated interactions governed by Zipfian aggregation processes naturally give rise to inverse power-law interaction forms. The next section formalizes this connection and derives the general structure of Zipfian interaction laws.

### 3. General framework for Zipfian interaction laws

The objective of this section is to establish a general mathematical framework in which inverse power-law interaction forms emerge naturally from Zipfian aggregation processes. Rather than postulating a specific functional dependence between interaction strength and distance, we derive the macroscopic interaction law from minimal statistical assumptions on microscopic heterogeneity and spatial aggregation. This section constitutes the core theoretical contribution of the paper.

### 3.1 Aggregated interaction fields

We consider a system composed of a large number of microscopic interaction sources distributed in space. Each source contributes locally to an effective interaction field, and the macroscopic interaction observed at a distance results from the aggregation of these contributions. Let $r$ denote the distance between an observation point and the sources of interaction.

We define the aggregated interaction field $I(r)$ as the sum or integral of microscopic contributions $i_j(r)$,

$$I(r) = \sum_j i_j(r),$$

or, in the continuum limit,

$$I(r) = \int \rho(\mathbf{x})\, i(\|\mathbf{x} - \mathbf{r}\|)\, d\mathbf{x},$$

where $\rho(\mathbf{x})$ denotes the spatial density of microscopic sources. At this stage, no assumption is made on the functional form of (i(r)), beyond positivity and monotonic decay with distance.

### 3.2 Microscopic heterogeneity

A central assumption of the framework is that microscopic interaction strengths are heterogeneous. Each microscopic contribution can be written as

$$i_j(r) = A_j\, f(r),$$

where $A_j$ is a random positive variable encoding the intrinsic strength of source $j$, and $f(r)$ is a distance-dependent attenuation function common to all sources. The variables $A_j$ are assumed to be drawn from a broad distribution without a characteristic scale, consistent with Zipfian aggregation.

This heterogeneity reflects the multiplicity of microscopic parameters influencing interactions in real systems, such as local intensities, capacities, propensities, or efficiencies. Importantly, the precise interpretation of $A_j$ is immaterial: only its statistical properties matter at the macroscopic level.

**3.3 Spatial aggregation over distance**

We now consider the cumulative effect of microscopic contributions within a spherical shell of radius (r) and thickness (dr). The number of sources contributing at distance (r) scales as

$$N(r) \sim r^{d-1} dr,$$

where (d) denotes the effective spatial dimension of the system. The aggregated interaction field can thus be expressed as

$$I(r) \sim \int A(r)\, f(r)\, r^{d-1}\, dr,$$

where $A(r)$ denotes the effective aggregated strength of sources at distance (r).

Under Zipfian aggregation, the cumulative contribution of heterogeneous microscopic sources within each shell is dominated by the largest terms rather than by the average. As a result, the effective strength $A(r)$ inherits scale-free statistics and does not introduce a characteristic length scale.

The statistical justification for extreme-value dominance and the resulting breakdown of self-averaging under Zipfian aggregation is provided in Appendix B.

**3.4 Micro-to-macro transition**

The transition from microscopic heterogeneity to macroscopic interaction laws follows from the absence of characteristic scales in both the distribution of $A_j$ and the spatial organization of sources. Dimensional consistency and scale invariance impose strong constraints on the functional form of $I(r)$.

Let us consider a rescaling of distances $r \to \lambda r$. Under such a transformation, the aggregated field must transform as

$$I(\lambda r) = \lambda^{-\alpha} I(r),$$

for some exponent $\alpha$. This homogeneity condition expresses the fact that no intrinsic scale is introduced by aggregation. The only functional form consistent with this requirement is a power law,

$$I(r) \sim r^{-\alpha}.$$

Thus, the emergence of inverse power-law interaction forms follows directly from scale invariance and aggregation, without the need to postulate any specific interaction law at the microscopic level.

These considerations lead to the following general proposition.

**Proposition 1 (Zipfian aggregation fixed point)**

Consider a spatially extended system of heterogeneous microscopic interaction sources whose intrinsic strengths follow a Zipf–Pareto distribution, and whose aggregation satisfies assumptions A1–A4 (heterogeneity, multiplicativity, scale invariance, and stability under coarse-graining).

Under Zipfian aggregation, the emergence of inverse power-law interaction fields is not an empirical regularity but a necessary consequence of scale invariance. In the absence of any intrinsic length scale, the aggregated interaction field admits a unique scale-invariant macroscopic form. In particular, the only admissible asymptotic scaling is a power-law decay,

$$I(r) \sim r^{-\alpha},$$

where the exponent $\alpha$ is an emergent quantity determined by effective dimensionality, symmetry, and aggregation structure, rather than by microscopic interaction details.

In isotropic systems embedded in an effective spatial dimension d, the inverse-square regime $\alpha = d - 1$ emerges as a stable statistical fixed point of the Zipfian aggregation process.

A formal derivation of this scaling result, based on shell decomposition, homogeneity arguments, and extreme-value dominance, is provided in Appendix A, with statistical justification discussed in Appendix B and extensions to effective dimensionality in Appendix C.

**3.5 Existence conditions for power-law interactions**

The derivation above relies on a number of minimal conditions. First, microscopic interaction strengths must be sufficiently heterogeneous to invalidate self-averaging and prevent convergence toward Gaussian statistics. The statistical conditions under which self-averaging breaks down in Zipfian regimes are discussed in Appendix B. Second, aggregation must be multiplicative or dominated by extreme contributions, as in Zipfian processes. Third, the spatial distribution of sources must be approximately homogeneous at large scales.

If these conditions are violated, deviations from pure power-law behavior may occur. For instance, finite-size effects, strong correlations, or the introduction of characteristic scales through external constraints can lead to truncated or crossover regimes. Nevertheless, within a broad range of conditions, inverse power-law interaction forms remain robust.

The value of the exponent $\alpha$ is not fixed by the general framework alone, but depends on symmetry, dimensionality, and aggregation structure. In particular, $\alpha$ reflects the balance between the growth in the number of contributing sources with distance and the decay of their individual influence.

In isotropic systems embedded in an effective spatial dimension $d$, simple dimensional arguments suggest $\alpha = d - 1 + \gamma$, where $\gamma$ encodes the scaling properties of microscopic contributions. The inverse-square law corresponds to a specific combination of dimensionality and aggregation symmetry, which will be analyzed in detail in the next section.

The important point is that $\alpha$ is an emergent quantity, not a parameter imposed externally. Different systems may exhibit different exponents while belonging to the same Zipfian universality class.

This section shows that inverse power-law interaction forms arise inevitably from Zipfian aggregation under minimal and generic assumptions. The next section demonstrates how the inverse-square law emerges as a privileged Zipfian fixed point under isotropic conditions.

## 4. The inverse-square law as a Zipfian fixed point

The objective of this section is to show that the inverse-square law, corresponding to an interaction exponent $\alpha = 2$, is not a special or ad hoc result, but a privileged fixed point of the Zipfian interaction framework under generic symmetry and dimensionality conditions. Rather than attributing the inverse-square law to specific physical mechanisms, we demonstrate that it emerges naturally from isotropic Zipfian aggregation in an effectively three-dimensional space.

### 4.1 Isotropy and symmetry constraints

We begin by considering the role of isotropy. An interaction field is said to be isotropic if its statistical properties are invariant under rotations of the underlying space. Isotropy is a minimal and natural assumption in the absence of directional bias or anisotropic constraints. Within the Zipfian aggregation framework developed in the previous sections, isotropy imposes strong restrictions on the admissible scaling behavior of the interaction field.

Under isotropic conditions, microscopic interaction sources contribute symmetrically in all spatial directions. As a result, the aggregated interaction field depends only on the radial distance *r* from the source, and not on angular coordinates. This symmetry ensures that the macroscopic interaction law is fully characterized by a single radial exponent.

Importantly, isotropy does not prescribe the value of the exponent $\alpha$ *a priori*. Rather, it constrains the form of the interaction field to be compatible with rotational invariance, thereby reducing the space of admissible scaling behaviors. The identification of $\alpha = 2$ as a fixed point follows from the combination of isotropy with dimensionality and aggregation arguments, as shown below.

**4.2 Effective spatial dimension**

The spatial dimension of the embedding space plays a central role in determining the scaling properties of aggregated interaction fields. Consider an isotropic space of effective dimension *d*. As shown in Section 3, the number of microscopic sources contributing within a spherical shell of radius *r* scales as $r^{d-1}$.

Under Zipfian aggregation, the cumulative contribution of sources within each shell is dominated by the largest contributions rather than by the mean. Consequently, the scaling of the interaction field reflects a balance between the geometric growth in the number of contributing sources and the decay of their individual influence with distance.

In the absence of additional characteristic scales, dimensional consistency requires that the interaction field scales inversely with the surface measure of the sphere in dimension *d*. This leads naturally to an interaction exponent of the form

$$\alpha = d - 1,$$

for isotropic Zipfian aggregation. In an effectively three-dimensional space $d = 3$, this condition yields $\alpha = 2$.

This result highlights the fact that the inverse-square law is a direct consequence of spatial dimensionality and isotropic aggregation, rather than a property tied to a particular physical force or interaction mechanism.

**4.3 Fixed-point interpretation**

The inverse-square law can be interpreted as a fixed point of the Zipfian interaction framework. Under coarse-graining transformations that preserve isotropy and scale invariance, interaction fields with exponents close to $\alpha = 2$ flow toward this value. Deviations from the inverse-square regime tend to be suppressed unless anisotropy, strong heterogeneity, or dimensional distortions are introduced.

From this perspective, the inverse-square law occupies a privileged position within the space of Zipfian interaction laws. It is not unique, but it is stable under a wide class of aggregation and renormalization processes. This stability explains both its ubiquity across disciplines and its robustness to microscopic variations.

### 4.4 Minimal geometric interpretation

The emergence of the inverse-square law admits a simple geometric interpretation. In isotropic three-dimensional space, the surface area of a sphere grows proportionally to $r^2$. When the influence of microscopic sources is distributed uniformly over this surface, the effective interaction strength per unit area decreases as the inverse of the surface measure. This purely geometric argument is consistent with, and subsumed by, the statistical framework developed here.

Crucially, this geometric interpretation should not be viewed as an alternative explanation competing with the Zipfian framework. Rather, it represents a particular manifestation of the same underlying principle: scale-invariant aggregation constrained by symmetry and dimensionality. The Zipfian perspective clarifies why such geometric arguments recur across otherwise unrelated systems.

### 4.5 Recovery of classical interaction laws

Within the present framework, several well-known interaction laws emerge as direct realizations of the same Zipfian fixed point. Newtonian gravitation and Coulomb electrostatics both exhibit inverse-square decay as a consequence of isotropy and three-dimensional embedding. These laws differ in their microscopic interpretation and coupling constants, but share an identical macroscopic scaling structure.

Similarly, gravity-type spatial interaction models used in geography and regional science can be understood within the same framework. In particular, Reilly-type formulations recover inverse-square distance decay under assumptions of isotropic spatial interaction and homogeneous embedding space. From the Zipfian perspective, these models do not borrow their structure from physics by analogy; rather, they instantiate the same statistical fixed point under socio-spatial aggregation.

This reinterpretation provides a unifying statistical explanation for the convergence of physical and spatial interaction laws toward the same functional form.

The relationship between the present statistical interpretation and classical field-theoretic derivations based on Gauss's law and conservative fields is briefly discussed in Appendix D.

### 4.6 Scope and limitations

While the inverse-square law emerges as a privileged Zipfian fixed point under isotropic and three-dimensional conditions, it should not be regarded as universal in all contexts. Anisotropic environments, network-mediated interactions, or effective dimensionalities different from three naturally lead to alternative scaling exponents, as will be discussed in the next section.

The present analysis nonetheless establishes that the prominence of the inverse-square law is neither accidental nor exclusively physical. It reflects a robust statistical and geometric fixed point of Zipfian aggregation processes.

This section demonstrates that the inverse-square law corresponds to a stable Zipfian fixed point arising from isotropy and effective dimensionality. The following section explores how relaxing these assumptions leads to generalized Zipfian interaction laws with exponents different from two.

Within this framework, deviations from the inverse-square law should not be interpreted as violations of physical principles, but as signatures of modified effective dimensionality, source heterogeneity, or aggregation structure.

## 5. Beyond the inverse-square law: generalized Zipfian exponents

The purpose of this section is to demonstrate that the Zipfian interaction framework developed in the previous sections does not merely reinterpret existing inverse-square laws, but naturally generates a broader family of interaction regimes characterized by exponents $\alpha \neq 2$. By relaxing the assumptions of isotropy, integer spatial dimensionality, or weak heterogeneity, the framework gives rise to generalized Zipfian interaction laws that extend beyond classical cases.

### 5.1 Effective non-integer dimensionality

The derivation of the inverse-square law in Section 4 relied on the assumption of an effectively three-dimensional isotropic embedding space. However, many real systems are more appropriately described by effective dimensionalities that differ from integer Euclidean dimensions. Examples include fractal spatial structures, constrained geometries, and systems embedded in heterogeneous or porous environments.

Let $d_{\text{eff}}$ denote the effective spatial dimension governing the growth of accessible interaction volume with distance. In such systems, the number of contributing sources within a shell of radius $r$ scales as

$$N(r) \sim r^{d_{\text{eff}} - 1}.$$

Under Zipfian aggregation, the balance between geometric growth and attenuation leads to an interaction exponent

$$\alpha = d_{\text{eff}} - 1.$$

When $d_{\text{eff}} \neq 3$, the resulting interaction law deviates naturally from the inverse-square form. This observation highlights that $\alpha = 2$ is not universal, but contingent on effective dimensionality.

Extensions to non-integer effective dimensions, fractal geometries, and network-mediated metrics are discussed in Appendix C.

**5.2 Strong heterogeneity and anomalous aggregation**

The general framework assumes heterogeneous microscopic contributions, but the degree of heterogeneity can vary significantly across systems. When heterogeneity is sufficiently strong, aggregation becomes dominated by extreme contributions rather than by geometric averaging. In such regimes, the effective scaling of the interaction field is modified.

Strong heterogeneity may arise from broad distributions of intrinsic strengths, spatial clustering of dominant sources, or correlations between source strength and location. Under these conditions, the effective exponent $\alpha$ reflects not only geometric constraints, but also the scaling properties of microscopic heterogeneity.

As a result, interaction laws of the form

$$I(r) \sim r^{-\alpha}, \alpha \neq d_{\text{eff}} - 1,$$

can emerge, even in nominally isotropic environments. These anomalous regimes define generalized Zipfian universality classes, in which extreme-value dominance reshapes macroscopic interaction decay..

**5.3 Anisotropy and network-mediated interactions**

Another important extension concerns anisotropic systems and network-mediated interactions. In many contexts, interactions do not propagate uniformly in all directions, but are constrained by underlying network structures, directional channels, or spatial heterogeneities. Examples include transportation networks, communication infrastructures, and social or economic interaction graphs.

In such systems, the effective interaction distance is not purely Euclidean, but reflects network topology and connectivity. The growth of accessible nodes with distance may follow non-Euclidean scaling laws, leading to modified interaction exponents. Within the Zipfian framework, these effects are captured by replacing Euclidean distance with an effective metric and by allowing directional dependence in aggregation.

Anisotropy and network constraints thus generate a continuum of interaction regimes characterized by distinct exponents $\alpha$, while preserving the underlying Zipfian aggregation mechanism.

**5.4 Regime classification and qualitative behavior**

The generalized Zipfian framework naturally leads to a qualitative classification of interaction regimes. In isotropic, weakly heterogeneous systems embedded in integer-dimensional space, the inverse-square law emerges as a stable fixed point. In contrast, systems with reduced effective dimensionality, strong heterogeneity, or anisotropic connectivity exhibit alternative scaling exponents.

These regimes should not be interpreted as violations of the Zipfian framework, but as different realizations of the same underlying principle under modified constraints. The value of $\alpha$, thus serves as an indicator of the effective geometry and aggregation structure of the system.

Importantly, the framework predicts smooth crossovers between regimes rather than abrupt transitions. As system properties evolve—through increasing heterogeneity, spatial restructuring, or network formation—the interaction exponent adjusts continuously, reflecting changes in effective dimensionality and aggregation dominance.

### 5.5 Implications for interaction modeling

The existence of generalized Zipfian exponents has important implications for interaction modeling across disciplines. Rather than calibrating interaction exponents empirically or borrowing canonical values by analogy, the present framework provides a principled basis for interpreting and predicting interaction decay.

In this view, observed deviations from inverse-square behavior are not anomalies requiring ad hoc explanations, but natural consequences of modified aggregation conditions. The Zipfian framework thus offers a unifying lens through which diverse interaction laws can be understood as members of a single statistical family.

In summary, this section demonstrates that the Zipfian interaction framework generates a rich spectrum of interaction laws extending beyond the inverse-square case. By varying effective dimensionality, heterogeneity, and anisotropy, a continuum of Zipfian exponents emerges. The following section presents a minimal illustrative model that exemplifies these mechanisms without relying on empirical data.

## 6. Illustrative example: a minimal Zipfian interaction model

The purpose of this section is to illustrate the Zipfian interaction framework through a minimal analytical model. The goal is not to reproduce any specific empirical system, but to demonstrate explicitly how inverse power-law interaction forms—and in particular the inverse-square regime—emerge from Zipfian aggregation under controlled and transparent assumptions. This illustrative example serves to clarify the mechanism without relying on real data or detailed system-specific modeling.

### 6.1 Model setup

We consider a simplified spatial system composed of a large number of microscopic interaction sources distributed uniformly in an isotropic three-dimensional space. Each source $j$ is characterized by an intrinsic interaction strength $A_j > 0$, drawn independently from a heavy-tailed distribution belonging to the Zipf–Pareto family. No characteristic scale is assumed for the distribution of $A_j$.

The contribution of a single source located at position $\mathbf{x}_j$ to the interaction field measured at position $\mathbf{r}$ is assumed to decay monotonically with Euclidean distance,

$$i_j(\mathbf{r}) = \frac{A_j}{g(\|\mathbf{r} - \mathbf{x}_j\|)},$$

where $g(r)$ is a positive attenuation function. At this stage, no specific functional form is imposed on $g(r)$.

The aggregated interaction field is defined as

$$I(\mathbf{r}) = \sum_j i_j(\mathbf{r}),$$

or equivalently, in the continuum limit,

$$I(r) = \int A(\mathbf{x}) \, g^{-1}(\|\mathbf{x} - \mathbf{r}\|) \, d\mathbf{x},$$

where isotropy ensures that the field depends only on the radial distance $r$.

**6.2 Zipfian aggregation over distance shells**

We now consider the contribution to the interaction field arising from sources located within a spherical shell of radius $r$ and thickness $dr$. The number of sources within this shell scales as

$$N(r) \sim r^2 \, dr,$$

reflecting the surface growth in three-dimensional space.

Under Zipfian aggregation, the cumulative contribution from each shell is dominated by the largest intrinsic strengths $A_j$ rather than by their average. As a consequence, the effective contribution of a shell at distance $r$ can be written as

$$\Delta I(r) \sim \frac{A_{\max}(r)}{g(r)} \, r^2 \, dr,$$

where $A_{\max}(r)$ denotes the dominant contribution within the shell. Due to the scale-free nature of the Zipfian distribution, $A_{\max}(r)$ does not introduce any characteristic length scale.

Dimensional consistency and scale invariance then require that the attenuation function $g(r)$ compensates the geometric growth of the shell in such a way that the aggregated field remains scale-free.

**6.3 Emergence of inverse power-law scaling**

Let us examine the scaling behavior of the aggregated interaction field under a rescaling of distance $r \to \lambda r$. In the absence of characteristic scales, the field must satisfy the homogeneity condition

$$I(\lambda r) = \lambda^{-\alpha} I(r),$$

for some exponent $\alpha$.

In three-dimensional isotropic space, balancing the growth of contributing sources ($r^2$) with attenuation leads naturally to

$$I(r) \sim r^{-2}.$$

This inverse-square scaling arises independently of the detailed form of $g(r)$, provided that the aggregation remains Zipfian and isotropic. The exponent $\alpha = 2$ thus appears as a stable outcome of the aggregation process rather than as an imposed assumption.

**6.4 Log–log representation and crossover behavior**

The behavior of the model can be conveniently visualized through a log–log representation of the aggregated interaction field. Plotting $\log I(r)$ as a function of $\log r$ reveals a linear regime with slope $-2$ over a broad range of distances, corresponding to the inverse-square fixed point identified in Section 4.

When deviations from isotropy or from strict Zipfian aggregation are introduced—for instance through finite-size effects or modified source distributions—the log–log plot exhibits smooth crossovers toward alternative slopes. These crossover regimes illustrate how generalized Zipfian exponents discussed in Section 5 emerge continuously as aggregation conditions are modified.

Importantly, no sharp transition is observed. The inverse-square regime appears as an attractor within the space of admissible scaling behaviors, consistent with its interpretation as a Zipfian fixed point.

**6.5 Interpretation and scope of the illustrative model**

This minimal model illustrates several key features of the Zipfian interaction framework. First, it shows explicitly that inverse power-law interaction forms can be derived without specifying detailed microscopic interaction laws. Second, it demonstrates that the inverse-square law arises naturally from isotropic Zipfian aggregation in three-dimensional space. Third, it highlights how alternative exponents emerge smoothly when aggregation conditions are altered.

The purpose of this example is not quantitative prediction, but conceptual clarification. Its simplicity allows the underlying mechanism to be identified transparently, without the confounding effects of empirical calibration or system-specific constraints. As such, it fulfills its role as an illustrative complement to the general theoretical framework.

The next section discusses the broader implications of these results and situates the Zipfian interaction framework within the wider context of statistical mechanics and complex systems.

## 7. Discussion

The framework developed in this paper proposes a statistical-mechanics interpretation of interaction laws traditionally regarded as domain-specific or physically fundamental. By framing inverse power-law interactions as emergent outcomes of Zipfian aggregation processes, the analysis shifts the focus from microscopic interaction mechanisms to universal properties of aggregation, symmetry, and scale invariance. This section discusses the implications, scope, and limitations of this perspective, as well as possible directions for future research.

### 7.1 Zipfian scaling as a unifying statistical principle

A central implication of the present work is the elevation of Zipfian scaling from an empirical regularity to a unifying statistical principle. In this interpretation, Zipf's law is not confined to rank–size distributions or specific growth models, but characterizes a broad universality class of aggregation processes dominated by heterogeneity and scale invariance [1,2,5]. When such processes are spatially mediated, they naturally generate long-range interaction fields governed by inverse power laws.

Zipfian scaling has been documented across systems of very different natures, including word frequencies, city sizes, firm distributions, and income hierarchies. Its recurrence has motivated extensive theoretical work showing that Zipf-type distributions can emerge from broad classes of stochastic and multiplicative processes under mild assumptions, suggesting that such scaling reflects fundamental probabilistic constraints rather than system-specific mechanisms [1,2,4].

This perspective provides a conceptual bridge between phenomena traditionally studied in isolation. Linguistic frequency distributions, urban hierarchies, economic concentrations, physical interaction fields, and spatial gravity models all exhibit power-law structures that can be interpreted as manifestations of the same underlying statistical mechanism. The recurrence of inverse-square and related interaction laws across disciplines thus reflects not a coincidence of physical analogies, but the robustness of Zipfian aggregation under minimal assumptions.

Paradoxically, this increasing microscopic complexity does not lead to increasingly complex macroscopic laws. On the contrary, when heterogeneous factors are aggregated under scale-invariant conditions, microscopic details are progressively washed out, and simple, universal interaction laws emerge. From this perspective, the inverse-square regime should not be interpreted as a signature of simplicity, but as the macroscopic imprint of highly complex and heterogeneous underlying processes.

**7.2 A statistical reading of physical interaction laws**

Within the proposed framework, classical physical laws such as Newtonian gravitation and Coulomb electrostatics can be reinterpreted as macroscopic realizations of Zipfian fixed points. This interpretation does not challenge their validity, predictive power, or physical foundations. Rather, it provides a complementary statistical reading that explains why similar functional forms emerge independently of the detailed nature of microscopic interactions.

From this viewpoint, inverse-square laws are not uniquely tied to specific force carriers or field equations, but arise because isotropic aggregation in three-dimensional space admits a stable scale-invariant fixed point. The physical derivations based on conservation laws and field equations remain essential for quantitative predictions and coupling constants, but the Zipfian framework clarifies why the same distance dependence recurs so systematically.

This statistical interpretation aligns naturally with broader developments in statistical physics, where macroscopic laws are increasingly understood as emergent properties insensitive to microscopic detail. The theory of universality and the renormalization group formalize how systems with very different microscopic interactions can exhibit identical scaling behavior at large scales [11–13]. In this sense, the present framework situates classical interaction laws within a wider landscape of universality rather than treating them as isolated principles.

**7.3 Interdisciplinary relevance and modeling implications**

The Zipfian interaction framework has direct implications for modeling practices across disciplines. In many applied contexts—such as geography, economics, transportation, and network science—interaction exponents are often calibrated empirically or borrowed from canonical models by analogy. The present analysis suggests that such exponents should instead be interpreted as emergent indicators of effective dimensionality, heterogeneity, and aggregation structure.

This shift has two important consequences. First, deviations from inverse-square behavior should not be viewed as anomalies requiring ad hoc explanations, but as signatures of modified aggregation conditions, such as anisotropy, network constraints, or strong heterogeneity. Second, the framework provides qualitative guidance for anticipating how interaction laws may evolve as system structure changes, for example through spatial reorganization or network formation.

More broadly, power-law scaling and heavy-tailed distributions have been documented across a wide range of physical, biological, and socio-economic systems, reinforcing the interdisciplinary relevance of Zipfian universality [1,5].

**7.4 Limitations of the framework**

Despite its generality, the proposed framework has clear limitations. First, it is inherently macroscopic and statistical in nature. It does not aim to replace detailed microscopic models, nor to derive coupling constants or system-specific parameters. Its explanatory power lies in functional form and scaling behavior, not in quantitative prediction.

Second, the framework assumes the absence of characteristic scales and relies on effective homogeneity at large distances. Systems dominated by strong boundaries, finite-size effects, or externally imposed scales may exhibit truncated or crossover behaviors not captured by pure Zipfian scaling. While such effects can often be incorporated through extensions, they lie outside the scope of the present analysis.

Third, the illustrative model presented here is deliberately minimal. While it clarifies the aggregation mechanism, it does not address dynamical aspects, temporal evolution, or feedback effects that may be essential in specific applications.

**7.5 Directions for future research**

Several directions naturally follow from the present work. A first avenue concerns the explicit treatment of anisotropic and network-mediated geometries, where effective distance metrics depart from Euclidean space and where interaction "range" is shaped by topology and accessibility constraints. In such settings, the key prediction of the framework is that the exponent $\alpha$ should track effective dimensionality and coarse-graining structure, suggesting a direct connection between observed scaling and the geometry induced by networks or constrained media [14,15].

A second direction involves dynamical Zipfian aggregation, where the distribution of source strengths and the effective interaction field co-evolve over time. This includes growth processes, adaptive networks, and non-equilibrium settings in which scale-free heterogeneity may arise endogenously and persist as a statistical attractor. Formalizing such dynamics would connect the present framework to a broader body of work on universality in stochastic processes and scaling in evolving complex systems [12,13].

Finally, empirical investigations could test a specific falsifiable implication of the framework: whether measured interaction exponents across real systems covary with independent indicators of (i) effective dimensionality (including fractal or constrained geometries), (ii) strength heterogeneity (tail index), and (iii) anisotropy/network constraints. Beyond calibration, such studies would help distinguish regimes where power-law interactions are true scaling fixed points from regimes dominated by finite-size truncation or crossovers—issues that require careful statistical assessment when dealing with heavy-tailed data [5].

### 7.6 Concluding perspective

Taken together, the results of this work suggest that Zipfian aggregation provides a coherent and unifying statistical foundation for understanding interaction laws across a wide range of systems. By reframing inverse power-law interactions as emergent properties of scale-invariant aggregation under heterogeneity and symmetry constraints, the framework clarifies both the ubiquity of classical interaction laws and the systematic origin of their observed variations.

From this perspective, inverse-square and related interaction laws should not be viewed as isolated consequences of specific physical mechanisms, but as robust statistical fixed points associated with particular symmetry and dimensionality conditions. This interpretation situates classical derivations within a broader universality framework, consistent with the central insights of statistical mechanics, according to which macroscopic laws are largely insensitive to microscopic detail and instead governed by collective constraints and invariance principles [11].

Beyond providing a reinterpretation of existing laws, the Zipfian framework opens a principled path toward the systematic classification and generalization of interaction regimes. By linking interaction exponents to effective dimensionality, aggregation structure, and heterogeneity, it offers a common language for physical, biological, and socio-spatial systems, and suggests that many apparent differences between domains reflect variations within a single universality class rather than fundamentally distinct mechanisms.

### 8. Conclusion

This paper has proposed a unified statistical-mechanics framework for understanding interaction laws that decay as inverse powers of distance. By interpreting such laws as emergent outcomes of Zipfian aggregation processes, we have shifted the focus from domain-specific microscopic mechanisms to generic properties of heterogeneity, scale invariance, symmetry, and spatial aggregation.

The core result of the analysis is that inverse power-law interaction forms arise inevitably under Zipfian aggregation when no characteristic scale is introduced. Within this framework, the inverse-square law is not a fundamental postulate nor a peculiarity of specific physical forces, but a privileged statistical fixed point associated with isotropic aggregation in an effectively three-dimensional space. Classical laws of gravitation and electrostatics, as well as gravity-type spatial interaction models, can thus be reinterpreted as particular realizations of the same Zipfian universality class.

Beyond the inverse-square regime, the framework naturally generates a continuum of generalized interaction laws when isotropy, effective dimensionality, or aggregation structure are modified. Variations in the interaction exponent emerge as systematic responses to changes in geometry, heterogeneity, or network constraints, rather than as anomalies or empirical adjustments. In this sense, the Zipfian perspective provides a principled explanation for both the robustness of classical interaction laws and the diversity of observed scaling behaviors across systems.

More broadly, the results suggest that Zipfian scaling should be viewed as a unifying statistical principle linking rank–size distributions, spatial hierarchies, and interaction fields within a common theoretical structure. The recurrence of inverse power-law interactions across physical, biological, and socio-spatial systems reflects the stability of scale-invariant aggregation mechanisms rather than the transfer of models by analogy between disciplines.

Future work may extend this framework toward explicitly dynamical settings, anisotropic and network-mediated geometries, or empirical investigations aimed at relating observed interaction exponents to independent measures of effective dimensionality and heterogeneity. By providing a coherent statistical foundation for interaction laws, the Zipfian aggregation framework opens a path toward a more unified understanding of scaling phenomena in complex systems.

These results suggest that Zipfian aggregation constitutes a natural statistical principle underlying the emergence and robustness of interaction laws in complex systems.

**Declaration of generative AI and AI-assisted technologies in the manuscript preparation process**

The author used AI-assisted tools for language editing and stylistic refinement. All scientific content, interpretations, and conclusions are solely those of the author.

## Appendix A. Formal derivation of the Zipfian interaction scaling

This appendix provides a formal derivation of the inverse power-law scaling of the aggregated interaction field,

$$I(r) \sim r^{-\alpha},$$

under the assumptions of Zipfian aggregation introduced in the main text. The purpose is to make explicit the mathematical steps underlying the micro-to-macro transition, without introducing additional conceptual hypotheses.

### A.1 General setup and assumptions

We consider a spatial system composed of microscopic interaction sources indexed by $j$, located at positions $\mathbf{x}_j \in \mathbb{R}^d$. Each source contributes to an interaction field measured at position $\mathbf{r}$. The microscopic contribution of source $j$ is written as

$$i_j(\mathbf{r}) = A_j\, f(\|\mathbf{r} - \mathbf{x}_j\|),$$

where:

- $A_j > 0$ denotes the intrinsic strength of source $j$,
- $f(r)$ is a deterministic attenuation function, positive and monotonically decreasing,
- $\|\cdot\|$ denotes the Euclidean norm.

The aggregated interaction field is defined as

$$I(\mathbf{r}) = \sum_j i_j(\mathbf{r}),$$

or, in the continuum limit,

$$I(\mathbf{r}) = \int_{\mathbb{R}^d} A(\mathbf{x})\, f(\|\mathbf{r} - \mathbf{x}\|)\, \rho(\mathbf{x})\, d\mathbf{x},$$

where $\rho(\mathbf{x})$ is the spatial density of sources.

The derivation relies on the following assumptions:

1. **Statistical homogeneity**:
   At large scales, the spatial distribution of sources is homogeneous and isotropic.
2. **Zipfian heterogeneity**:
   The intrinsic strengths $A_j$ are independent realizations of a heavy-tailed distribution belonging to the Zipf–Pareto family, with no characteristic scale.

3. **Scale invariance**:
   No external length scale is introduced by the aggregation process.
4. **Dominance of extreme contributions**:
   Due to Zipfian heterogeneity, aggregation is non-self-averaging and dominated by extreme values rather than by means [16,17].

### A.2 Shell decomposition and aggregation

We decompose space into spherical shells centered at $\mathbf{r}$, of radius $s = \|\mathbf{x} - \mathbf{r}\|$ and thickness $ds$. In $d$ dimensions, the volume of such a shell scales as

$$dV(s) \sim s^{d-1}\, ds.$$

The contribution of a shell at distance $s$ to the interaction field is

$$dI(s) \sim \sum_{j \in \text{shell}(s)} A_j\, f(s).$$

Under Zipfian aggregation, the sum over sources within a shell is dominated by the largest intrinsic strength $A_{\max}(s)$ in that shell. Hence,

$$dI(s) \sim A_{\max}(s)\, f(s)\, s^{d-1}\, ds.$$

Because the distribution of $A_j$ is scale-free, the statistical properties of $A_{\max}(s)$ depend only on the number of contributing sources and do not introduce an intrinsic length scale. As a result, $A_{\max}(s)$ contributes a multiplicative factor that preserves scale invariance.

### A.3 Homogeneity and scaling constraint

We now examine the behavior of the interaction field under a rescaling of distances,

$$\mathbf{r} \to \lambda \mathbf{r}.$$

Scale invariance implies that the aggregated field transforms as a homogeneous function of degree $-\alpha$,

$$I(\lambda r) = \lambda^{-\alpha} I(r),$$

for some exponent $\alpha > 0$.

This functional equation admits only power-law solutions. Hence, independently of the detailed form of $f(r)$,

$$I(r) \sim r^{-\alpha}.$$

The value of $\alpha$ is determined by the balance between:

- geometric growth of contributing sources ($s^{d-1}$),
- attenuation through $f(s)$,
- and the statistical dominance of extreme values induced by Zipfian heterogeneity.

### A.4 Conditions of existence and convergence

The above derivation holds provided that the following conditions are satisfied:

- The effective aggregation over shells converges in the infrared and ultraviolet limits.
- The tail exponent of the Zipf–Pareto distribution is such that extreme-value dominance persists at all relevant scales.
- Spatial correlations between $A_j$ and $\mathbf{x}_j$ do not introduce a characteristic length scale.

If these conditions are violated, truncated or crossover regimes may emerge, leading to deviations from pure power-law behavior. These cases are discussed qualitatively in Section 5.

### A.5 Summary

This appendix establishes formally that inverse power-law interaction fields arise inevitably from Zipfian aggregation under minimal and generic assumptions. The scaling

$$I(r) \sim r^{-\alpha}$$

is not postulated, but follows from homogeneity, heavy-tailed heterogeneity, and spatial aggregation. The main text focuses on the interpretation and implications of this result, while the present appendix provides its mathematical foundation.

**Appendix B. Extreme-value dominance under Zipfian aggregation**

This appendix provides a statistical justification for the dominance of extreme contributions in Zipfian aggregation processes. This property underlies the non-self-averaging behavior assumed throughout the paper and explains why aggregated interaction fields are governed by scale-free statistics rather than by mean-field averages.

**B.1 Pareto distributions and heavy-tailed heterogeneity**

Let $\{A_j\}_{j=1}^{N}$ be a collection of independent positive random variables drawn from a Pareto-type distribution,

$$P(A > a) \sim a^{-\mu},$$

with tail exponent $\mu > 0$. Such distributions are characterized by heavy tails and the absence of a characteristic scale [3,1].

For $\mu \leq 2$, the variance of $A$ diverges, and for $\mu \leq 1$, even the mean is undefined. In these regimes, classical self-averaging arguments based on the law of large numbers fail. As a consequence, sums of such variables exhibit large sample-to-sample fluctuations and are not well described by their expected value.

**B.2 Order statistics and dominance of extremes**

The statistical behavior of the sum

$$S_N = \sum_{j=1}^{N} A_j$$

can be analyzed using order statistics. Let $A_{(1)} \geq A_{(2)} \geq \cdots \geq A_{(N)}$ denote the ordered values of the sample.

For Pareto-distributed variables, it is well known that the maximum value scales as

$$A_{(1)} \sim N^{1/\mu},$$

while the contribution of lower-ranked terms decays rapidly [16,18]. As a result, for sufficiently heavy tails, the sum $S_N$ is asymptotically dominated by a finite number of largest terms,

$$S_N \approx A_{(1)} + A_{(2)} + \cdots$$

This extreme-value dominance implies that aggregation does not lead to Gaussian statistics, even for large $N$. Instead, macroscopic quantities inherit the scale-free properties of the underlying distribution.

### B.3 Non-self-averaging regimes

A key consequence of extreme-value dominance is the breakdown of self-averaging. In self-averaging systems, relative fluctuations vanish in the thermodynamic limit. By contrast, in Zipfian regimes,

$$\frac{\mathrm{Var}(S_N)}{\mathbb{E}[S_N]^2} \not\to 0 \text{ as } N \to \infty.$$

Such non-self-averaging behavior is well documented in statistical physics, particularly in systems with quenched disorder and heavy-tailed distributions [19,20]. In these systems, macroscopic observables remain sensitive to microscopic extremes even at large scales.

Within the context of spatial aggregation, this implies that the contribution of a shell of interaction sources is controlled by its most influential elements rather than by an average density. This property is essential for the emergence of Zipfian interaction fields.

This behavior is characteristic of non-self-averaging phases in disordered systems.

### B.4 Implications for spatial aggregation

When Zipfian-distributed interaction strengths are aggregated over space, the dominance of extreme values ensures that no characteristic scale is introduced by averaging. As the number of contributing sources increases with distance, the effective interaction strength remains governed by scale-invariant statistics.

This mechanism provides the statistical foundation for the results derived in Sections 3–5. It explains why inverse power-law interaction forms persist under aggregation and why their scaling exponents are insensitive to microscopic details, depending instead on geometric and symmetry constraints.

### B.5 Summary

This appendix establishes that Zipfian aggregation processes are inherently dominated by extreme contributions, as a direct consequence of Pareto-type heavy tails and order-statistics effects. The resulting non-self-averaging behavior justifies treating macroscopic interaction fields as scale-free objects governed by universality rather than by mean-field averages. This statistical property is central to the Zipfian interpretation of interaction laws developed in the main text.

**Appendix C. Alternative geometries and effective dimensions**

This appendix discusses extensions of the Zipfian interaction framework to alternative geometries and effective dimensions. These cases complement the results presented in Section 5 and illustrate how deviations from the inverse-square law arise naturally when the assumptions of isotropy, Euclidean embedding, or integer dimensionality are relaxed. No new hypotheses are introduced; the purpose is to clarify the scope of applicability of the framework.

**C.1 Effective dimensions different from three**

The derivation of the inverse-square law as a Zipfian fixed point relies on an effective spatial dimension $d = 3$. However, many systems of interest are embedded in environments where the effective dimension governing interaction growth differs from the ambient Euclidean dimension.

In such cases, the number of accessible interaction sources within a shell of radius $r$ scales as

$$N(r) \sim r^{d_{\text{eff}}-1},$$

where $d_{\text{eff}}$ denotes the effective dimension. This concept is well established in the study of disordered and constrained systems, where transport and interaction properties are controlled by reduced or anomalous dimensionality [21,22].

Within the Zipfian aggregation framework, replacing $d$ by $d_{\text{eff}}$ directly leads to generalized interaction exponents,

$$\alpha = d_{\text{eff}} - 1,$$

consistent with the results discussed in Section 5.

**C.2 Fractional and fractal geometries**

A particularly important class of systems is characterized by fractal or self-similar spatial organization. In such systems, the effective dimension is non-integer and reflects the scaling properties of the underlying structure rather than its embedding space.

Fractal dimensions arise naturally in porous media, percolation clusters, urban morphology, and complex spatial networks. Their role in governing transport, diffusion, and interaction processes has been extensively studied in statistical physics [23,24].

When Zipfian aggregation occurs on a fractal support of dimension $d_f$, the interaction field inherits this non-integer geometry. The resulting interaction law follows

$$I(r) \sim r^{-(d_f-1)},$$

up to logarithmic corrections in marginal cases. This extension illustrates that deviations from inverse-square behavior can be interpreted as direct signatures of fractal spatial organization rather than as failures of the interaction model.

**C.3 Effective metrics and network-mediated distances**

In many systems, interactions are not governed by Euclidean distance but by effective metrics defined by network topology, accessibility, or transport cost. Examples include transportation networks, communication infrastructures, and social or economic interaction graphs.

In such contexts, the growth of accessible nodes with distance is controlled by network properties such as degree distribution, shortest-path scaling, and clustering. These effects can be captured by replacing the Euclidean distance $r$ with an effective distance $r_{\text{eff}}$, leading to modified scaling relations [14,25].

Within the Zipfian framework, aggregation over network-mediated distances preserves the dominance of extreme contributions but modifies the effective dimensionality. The resulting interaction laws remain of inverse power-law form, with exponents reflecting network geometry rather than physical space.

**C.4 Relation to generalized interaction regimes**

The cases discussed above demonstrate that generalized Zipfian exponents arise naturally from geometric and metric considerations, without altering the underlying aggregation mechanism. Whether due to reduced dimensionality, fractal structure, or network constraints, deviations from $\alpha = 2$ correspond to changes in effective geometry rather than to ad hoc modifications of interaction rules.

This observation reinforces the interpretation proposed in the main text: inverse power-law interaction laws form a continuous family governed by Zipfian universality, with the inverse-square law representing a particularly symmetric and stable case.

**C.5 Summary**

This appendix shows that the Zipfian interaction framework extends naturally to alternative geometries and effective dimensions. Fractional dimensionality, constrained spatial supports, and network-mediated metrics all lead to generalized interaction exponents while preserving the core statistical mechanism of Zipfian aggregation. These extensions confirm that the framework is not restricted to classical Euclidean settings and provide a coherent interpretation of diverse interaction regimes discussed in Section 5.

**Appendix D. Relation to classical derivations**

This appendix briefly situates the Zipfian interaction framework with respect to classical physical derivations of inverse-square laws. The purpose is not to reproduce established physical arguments, but to clarify the conceptual relationship between field-theoretic approaches and the statistical interpretation proposed in the main text.

**D.1 Relation to Gauss's law and geometric arguments**

In classical physics, inverse-square laws arise naturally from Gauss's law and related geometric considerations. For conservative fields generated by point sources in isotropic three-dimensional space, flux conservation implies that field intensity decreases proportionally to the inverse of the surface area of a sphere, leading directly to a $1/r^2$ dependence [26].

This derivation relies on two key ingredients: isotropy and three-dimensional embedding. These same ingredients appear in the Zipfian framework, but play a different conceptual role. Rather than enforcing flux conservation through field equations, isotropy and dimensionality enter as statistical constraints governing aggregation over space [26,27].

From the Zipfian perspective, the inverse-square law emerges because isotropic aggregation in three-dimensional space admits a stable scale-invariant fixed point. The geometric argument based on surface growth is thus recovered as a particular manifestation of a more general statistical mechanism, rather than as an independent principle.

**D.2 Conservative fields and potential-based formulations**

Classical inverse-square interaction laws are typically formulated in terms of conservative fields derived from scalar potentials satisfying Laplace or Poisson equations. These formulations are essential for describing field dynamics, boundary conditions, and coupling to sources [28].

The present framework does not seek to replace or reinterpret these field-theoretic constructions. Instead, it addresses a complementary question: why do effective interaction fields, once aggregated over heterogeneous sources, so often exhibit inverse power-law decay independently of their microscopic origin?

In this sense, the Zipfian approach operates at a different explanatory level. It accounts for the functional form of interaction decay without specifying the underlying field equations, coupling constants, or conservation laws. Classical derivations remain indispensable for quantitative modeling, while the Zipfian framework explains the robustness and recurrence of the resulting scaling behavior.

**D.3 Complementarity rather than competition**

The statistical interpretation proposed in this paper should therefore be understood as complementary to classical physical derivations. It neither contradicts nor supersedes field-theoretic approaches, but situates them within a broader landscape of universality.

From this viewpoint, Newtonian gravitation and Coulomb electrostatics correspond to specific realizations of a Zipfian fixed point under strong symmetry and conservation constraints. Other systems—physical, biological, or socio-spatial—may realize the same fixed point without sharing the same microscopic equations, provided that aggregation, heterogeneity, and symmetry conditions are satisfied.

**D.4 Summary**

This appendix clarifies that the Zipfian framework is consistent with classical geometric and field-theoretic derivations of inverse-square laws. While traditional approaches derive interaction laws from conservation principles and field equations, the Zipfian interpretation explains their ubiquity as a consequence of scale-invariant aggregation under isotropy and dimensionality constraints. The two perspectives address different questions and operate at complementary levels of description.